\documentclass[aps,onecolumn]{revtex4}
\usepackage{amsfonts}
\usepackage{amsmath}
\usepackage{amsopn}
\usepackage{amssymb}
\usepackage{graphicx}
\usepackage{graphics}
\usepackage{color}
\usepackage{verbatim}

\begin{document}

\title{Commentary to article \textit{Spatiotemporal optical solitons}, by B.
A. Malomed, D. Mihalache, F. Wise, and L.Torner, J. Opt. B: Quantum
Semiclass. Opt. \textbf{7}, R53-R72 (2005)}
\author{}
\maketitle


This review article (\textbf{RA}) was a brief but comprehensive survey of
the general area of \textit{multidimensional solitons}, i.e., two- and
three-dimensional (2D and 3D) modes self-trapped as a result of the
competition between the linear effects of diffraction and dispersion, which
tend to stretch any wave packet in spatial and temporal directions, and
nonlinear self-compression of the wave field. The RA has produced an
appreciable impact, having been cited 526 times (Google Scholar, as of Febr.
12, 2016).

Multidimensional solitons draw continuously renewed interest in many
branches of physics, finding especially important realizations in nonlinear
photonics (optics and plasmonics) and Bose-Einstein condensates (BECs); in
particular, spatiotemporal optical solitons are also known as ``light
bullets". The solitons are classified as \textit{fundamental} ones, which
carry no topological structure, and various topological modes, including 2D
and 3D solitons with embedded vorticity, and more sophisticated 3D states,
such as \textit{hopfions}, i.e., vortex tori with intrinsic twist, which
carry two independent topological numbers.

Unlike 1D solitons, which are normally stable, their multidimensional
counterparts are vulnerable to severe instabilities. Indeed, the ubiquitous
cubic self-attractive nonlinearity, which readily creates solitons,
simultaneously gives rise to the critical and supercritical \textit{collapse}%
, i.e., spontaneous formation of a singularity after a finite propagation
distance or time, in the 2D and 3D geometries, respectively. The collapse
destabilizes fundamental solitons, while their vortex solitons are subject
to a still stronger splitting instability against perturbations breaking the
axial symmetry of the vortices. Accordingly, a challenging problem is search
for physically relevant settings which admit stabilization of the solitons,
the settings being categorized according to the underlying stabilization
mechanisms. Several generic mechanisms have been identified in the RA and
developed in subsequent works: (i) the use of quadratic
(second-harmonic-generating) or saturable nonlinearities, which do not lead
to the collapse, and thus make fundamental solitons automatically stable,
but failing to stabilize vortices; (ii) effective trapping potentials (in
particular, spatially periodic ones, such as photonic lattices in optics, or
optical lattices in BEC), which may stabilize 2D and 3D solitons of all
types (lattice potentials create stable vortex solitons in the form of
multipeak complexes, with the vorticity represented by phase shifts between
adjacent peaks); (iii) competing nonlinearities, such as combinations of
attractive cubic and repulsive quintic terms, which secure partial
stabilization of vortices; (iv) \textit{management} techniques, which impose
periodic switch of the nonlinearity between attraction and repulsion, making
it possible to stabilize 2D fundamental solitons. Following the review
provided in the RA, these techniques, as well as their combinations, have
been developed in a large number of works, revealing many possibilities for
the creation of stable solitons of different types.

An essential peculiarity of this research area, which was stressed in the
RA, and remains obvious presently, is disbalance between a very large number
of theoretical predictions and few experimental results. Nevertheless, in
the course of 10 years since the publication of the RA, several essential
experimental findings have been published. These include, in particular, a
soliton in the\ BEC of $^{85}$Rb atoms with aspect ratio $2.5$ of the
trapping potential, which makes the soliton's shape close to isotropic \cite%
{Rb-85}; fundamental \cite{array1} and vortex \cite{array2}
\textquotedblleft optical bullets" created in arrays of optical fibers,
which may be considered as semi-discrete media; the creation of (2+1)D
spatial fundamental solitons in bulk optical media with a cubic-quintic \cite%
{CQ} 
competing nonlinearity; 2D exciton-polariton \textit{gap solitons}
(supported by a lattice potential) in a microcavity \cite{plasmon};
direct observation of filamentation of ultrashort laser pulses in a Kerr
medium in the case of negative group-velocity dispersion (which is necessary
for the formation of \textquotedblleft bullets") \cite{filamentation}, and
the observation of a characteristic structure of the self-compressing
\textquotedblleft bullet", composed of a high-density core and a surrounding
ring pattern \cite{core}. The creation of truly stable multidimensional
vortex solitons in continuous media has not been reported yet (solitary
vortices, whose limited stabilization is supported by nonlinear loss, were
observed very recently \cite{vortex}).

The theoretical work in the area has been developing in many directions
since the publication of the RA, being, to a large extent, stimulated by
results summarized in it. In particular, completely new settings allowing
the creation of multidimensional solitons have been elaborated. One of them
is the use of $D$-dimensional media with \emph{repulsive} nonlinearity,
whose local strength grows from the center to periphery, as a function of
distance $r$, at any rate faster than $r^{D}$. This setting supports a
variety of robust 2D and 3D solitons, including quite sophisticated ones,
such as \textit{soliton gyroscopes} and hopfions \cite{ICFO}. This direction
is related to a still broader area of studies of solitons in media with
effective potentials induced by spatial modulation of the nonlinearity \cite%
{RMP}. Further, recent considerations of BEC with linear \textit{spin-orbit
coupling} between its components give rise to objects which were assumed
impossible: \emph{stable solitons} in free 2D \cite{HS} and 3D \cite{HP}
space, supported by the attractive cubic nonlinearity, without the help of
any trapping potential. They are built of mixed fundamental and vortical
components, see the figure. These findings help to understand the profound
difference between the stabilization in 2D and 3D settings: in the former
case, the system creates stable solitons as \textit{ground states}, while in
3D a ground state cannot exist in the presence of the supercritical
collapse, the solitons being metastable modes.
\begin{figure}[tbh]
\centering\includegraphics[width=0.50\columnwidth]{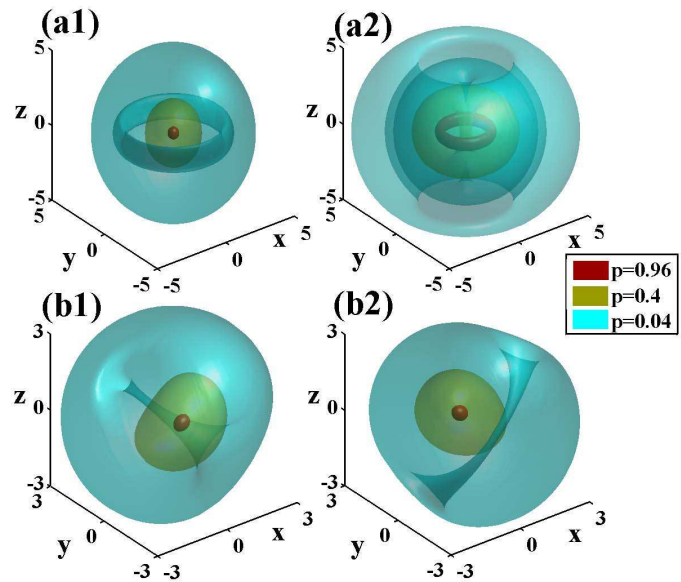}
\caption{(Color online) Examples of metastable three-dimensional solitons
produced by the two-component BEC model with spin-orbit coupling, as per
Ref. \protect\cite{HP}: (a1,a2) density (\textbf{p}) profiles of the
fundamental and vortex components; (b1,b2) the same in a soliton with the
fundamental and vortex terms mixed in each component.}
\end{figure}

\end{document}